\documentclass[conference]{IEEEtran}

\usepackage{cite}
\usepackage{amsmath,amssymb,amsfonts, amsthm}
\usepackage[table]{xcolor}
\usepackage{graphicx}
\usepackage{textcomp}
\usepackage{graphicx}
\usepackage{multirow}
\usepackage{subcaption}
\usepackage[ruled,linesnumbered, noend]{algorithm2e}
\usepackage{algpseudocode}
\usepackage{comment}
\usepackage{multicol}
\usepackage{vwcol}
\usepackage{tikz}
\usepackage[most]{tcolorbox}
\usepackage{todonotes}
\usepackage{listings}
\usepackage{wrapfig}
\usepackage[hidelinks]{hyperref}
\usepackage{url}
\usepackage{lipsum}

\usepackage{floatrow}
\newfloatcommand{capbtabbox}{table}[][\FBwidth]
\usepackage{caption}

\usetikzlibrary{fit}
\usetikzlibrary{patterns}

\usepackage{soul}
\usepackage{booktabs}

\newcommand{\redcite}[1]{{\textcolor{red}{\cite{#1}}}}

\usepackage{enumitem}
\begin{document}

\title{On the Partitioning of GPU Power among Multi-Instances}

\author{\IEEEauthorblockN{Tirth Vamja}
	\IEEEauthorblockA{\textit{IBM Research India and IIT Madras} \\
		}
	
	
	\and
	\IEEEauthorblockN{Kaustabha Ray}
	\IEEEauthorblockA{\textit{IBM Research - India}
		}
		
	\and
	\IEEEauthorblockN{Felix George}
	\IEEEauthorblockA{\textit{IBM Research - India}
	}
	
	\and
	\IEEEauthorblockN{UmaMaheswari C Devi}
	\IEEEauthorblockA{\textit{IBM Research - India}
	}
}


\maketitle
\pagestyle{plain}
\pagenumbering{arabic}

\begin{abstract}

Efficient power management in cloud data centers is essential for reducing costs, enhancing performance, and minimizing environmental impact. GPUs, critical for tasks like machine learning (ML) and GenAI, are major contributors to power consumption. NVIDIA’s Multi-Instance GPU (MIG) technology improves GPU utilization by enabling isolated partitions with per-partition resource tracking, facilitating GPU sharing by multiple tenants. However, accurately apportioning GPU power consumption among MIG instances, which is required for proper energy and carbon accounting and management, remains challenging due to  lack of hardware support.
This paper addresses this challenge by developing software methods to estimate power usage per MIG partition. We analyze NVIDIA GPU utilization and power consumption metrics and find that light-weight methods with good accuracy can be difficult to construct. We hence explore the use of  ML-based power models to enable accurate, partition-level power estimation. Our findings reveal that a single generic offline power model or modeling method is not applicable across diverse workloads, especially with concurrent MIG usage, and that workload-specific models constructed using partition-level utilization metrics of workloads under execution can significantly improve accuracy. We also propose scaling the MIG power predicted by the models using total measured GPU power, when available, to eliminate  aggregate error and reduce error in per-MIG estimations. Using NVIDIA A100 GPUs, we demonstrate and evaluate our approach for
accurate partition-level power estimation for workloads including matrix multiplication and Large Language Model inference, contributing to transparent and fair carbon reporting. 
\end{abstract}


\IEEEpeerreviewmaketitle

\section{Introduction}
\label{sec:intro}
\noindent
Cloud data centers have become the backbone of modern computing, enabling a wide spectrum of applications and services, from social media platforms to complex scientific computations in a cost-efficient manner\redcite{li2023btvmp, wang2024designing}. As data centers scale, efficient power management is increasingly critical to curb rising energy costs and meet global carbon emission regulations \redcite{ghg-reporting-1,ghg-reporting-2}. It also helps reduce operational overhead, improve performance, and lower environmental impact. Among the various components of a data center, Graphics Processing Units (GPUs) play a critical role, particularly in high-performance computing tasks such as machine learning, data analytics, and scientific simulations. With the increasing adoption of GenAI-based applications, 
the projected growth for data center GPU's during 2023-28 is 
 34.6\%\redcite{dc-gpu-growth}, necessitating the development of efficient strategies for improving their utilization levels and energy efficiency.

The Multi-Instance GPU (MIG) technology of NVIDIA available from its A100 data center GPUs of the Ampere series is one significant attempt at advanced GPU resource management\redcite{NVIDIA-MIG, li2022miso, MIG-char}.
MIG enables virtualization and partitioning of a single GPU into multiple isolated instances, referred to as {\em partitions\/}, allowing multiple tenants to share a single GPU without interference or performance degradation. This is particularly advantageous in cloud environments, where strict isolation is necessary to guarantee data as well as performance security to persuade users to share resources with other cross-organizational users. Such sharing can ameliorate GPU under-utilization and expand access to this scarce and expensive resource to a larger number of users.

To be truly useful, the increased flexibility offered by MIG should be accompanied by systems for accurately tracking the consumption of each partition and user of the GPU for various costing and reporting purposes. While such accounting can be seamless for the GPU's compute and memory units, the same cannot be said about the power it consumes.
This is because compute and memory cycles of each process can be separately tracked with hardware and firmware support, but not power.  Nevertheless, measuring power consumption at the partition-level is also crucial because GPU's are much more power-hungry than CPU's and the power drawn is a major source of green-house gas (GHG) emissions in the data center. Given the increasing thrust on reducing the GHG emissions due to computing and mandatory emission reports demanded by the regulatory bodies, cloud operators are obligated to determine each partition's fair share of the aggregate GPU power.

In this paper, we explore approaches to perform the aforementioned partitioning using appropriate attributes of the GPU as determinants. We set out to devise some simple,  lightweight approaches that can be implemented in a low-level module, preferably the GPU's device driver, to enable easy, real-time access to power per MIG instance. Such a lightweight approach will be contingent at minimum on the following factors: (i) easy-to-observe GPU runtime parameters that drive power consumption, (ii) consistency of GPU's power performance characteristics across workloads and  (iii) additivity of power consumption for co-located workloads.
To determine whether these factors hold, we perform extensive benchmarking using NVIDIA's Volta V100\redcite{NVIDIA-V100} and Ampere A100\redcite{NVIDIA-A100}  GPUs and analyze the captured multi-dimensional data extensively, the results of which rule out the possibility of a lightweight approach, at least for the GPUs considered. Nevertheless, our analytical approach can be applied for other classes of GPU's in which power partitioning is required.

We next consider increasingly complex methods for our task at hand. Our first approach is to construct a universal power model that is applicable for a variety of common workloads and use it for independently estimating each partition's power. It turns out that  with ML methods such as Extreme Gradient Boosting (XGBoost)\redcite{XGBoost}, errors can be high with unified models constructed using data sets from dissimilar multiple workloads, in non-MIG, full GPU mode. However, for unified models constructed from similar  workloads, the error is substantially reduced.  This highlights the potential advantages of applying workload-aware models, which we examine further. Interestingly, we demonstrate that individual workload-aware models do not significantly reduce prediction error when applied to MIG power prediction, whilst a unified model adds to predictivity.
Finally, as a measure of improving accuracy, with all the methods, we scale the model-estimated MIG power values using aggregate measured GPU power, when available, and evaluate its efficacy.

To summarize, our high-level contributions include:
\begin{itemize}
    \item Utilization vs. Power characterization for NVIDIA V100 and A100 GPUs via extensive benchmarking.
    \item Construction and evaluation of GPU power models using lightweight metrics.
    \item Highlighting the challenges posed by the problem of partitioning GPU power among MIG instances and showing that lightweight approaches that can be implemented in a low-level module may not be possible.
    \item Design and evaluation of ML-based approaches to partition GPU power among MIG instances and a scaling method to lower errors.
\end{itemize}

The rest of this paper is organized as follows:  Sec.~\ref{sec:background} presents an overview of NVIDIA GPU Architecture. 
Sec.~\ref{sec:detailed} presents our benchmarking results and 
discusses the software-hardware co-implications  for modeling GPU power consumption and estimation towards GPU partitions. Sec.~\ref{sec:GPU-power-partitioning-MIG} proposes and evaluates approaches for dividing full GPU power across MIG partitions. Sec.~\ref{sec:rel}  discusses related work, while Sec.~\ref{sec:conc} concludes the paper and indicates some future directions.
\section{Overview of NVIDIA GPU Architecture} \label{sec:background}
Graphics Processing Units (GPUs), originally intended for accelerating computer graphics\redcite{GPU-evolution}, have emerged as a cornerstone of high-performance computing 
due to their highly-parallel processor architecture, consisting of hundreds to thousands of small cores.
In NVIDIA GPUs\footnote{Unless otherwise specified, GPU henceforth refers to NVIDIA's GPUs.}, the processing cores are organized into Streaming Multiprocessors (SMs)\redcite{NVIDIA-GPUs, NVIDIA-VOLTA, NVIDIA-A100, NVIDIA-HOPPER}. 
Each SM  consists of two types of cores, {\em CUDA\/}\redcite{NVIDIA-CUDA-Cores} and {\em Tensor\/} cores\redcite{NVIDIA-Tensor-Cores, NVIDIA-Tensor-Cores-Prog}, to support arithmetic operations of different types and precisions, such as 16-, 32-, and 64-bit floating points (FP16, FP32, and FP64 resp.) and 8- and 16-bit integers (INT8 and INT16, resp.). CUDA cores handle scalar operations per cycle, while Tensor cores accelerate AI workloads with mixed-precision matrix operations. Due to their simpler design, CUDA cores far outnumber Tensor cores in an SM.
GPUs feature sophisticated L1/L2 caches, shared memory, and high-bandwidth off-chip memory, enhancing performance in data-intensive tasks.

The above hardware support for parallelism is accompanied by a matching parallel software programming and execution model based on {\em warps\/}.
A warp, typically a group of 32 threads, executes the same instruction on different data within an SM, which schedules and manages their execution. If a warp stalls, the SM switches to a ready warp, minimizing latency and keeping cores fully utilized. This fine-grained multithreading, following the SIMT/SIMD model, maximizes throughput by exploiting data-level parallelism in workloads like AI.


GPU performance is assessed through hardware metrics such as core clock speed (SMCLK), memory utilization (DRAMA), CUDA core (FP64A, FP32A, FP16A, INT16A) and tensor core (TENSO) usage, and power consumption (POWER) \redcite{NVIDIA-DCGM-Metrics}. These metrics help optimize performance and energy efficiency. In data centers, they can be monitored with minimal overhead using NVIDIA's Data Center GPU Manager (DCGM) \cite{NVIDIA-DCGM}.


NVIDIA Nsight Compute\redcite{NVIDIA-Nsight-Compute} offers detailed, kernel-level GPU profiling, analyzing memory access, warp execution, and thread divergence. While valuable for workload optimization, it incurs higher overhead and is suited for development rather than real-time monitoring. This study evaluates both DCGM and Nsight metrics for power prediction, balancing detail and efficiency for cloud data centers.

\begin{wraptable}{r}{0.21\textwidth}
\vspace{-0.1cm}
    \centering
    {
    \begin{tabular}{|c|c|c|}
    \hline
    \textbf{Profile} & \textbf{Mem.} & \textbf{SM } \\ 
    {} &              \textbf{Frac.} & \textbf{Frac.} \\ \hline
    1g.10gb & 1/8  & 1/7 \\ \hline
    1g.20gb & 2/8 & 1/7 \\ \hline
    2g.20gb & 2/8 & 2/7 \\ \hline
    3g.40gb & 4/8 & 3/7 \\ \hline
    4g.40gb & 4/8 & 4/7 \\ \hline
    7g.80gb & 8/8 & 7/7 \\ \hline
    \end{tabular}
    }
    \vspace{-0.2cm}
    \caption{\small GPU Partition Profiles on A100-80GB}
    \label{tab:MIG-profiles}
\end{wraptable}
\vspace{-0.1cm}
\emph{\hspace{-0.22cm}Multi-Instance GPU (MIG):} 
Modern data center GPUs offer up to a thousand TFLOPS\redcite{NVIDIA-A100-tput}, 
but workloads may not be capable of fully utilizing this capacity. To improve the overall utilization, extend access to wider set of users, and also reduce cost, different sharing modes namely, time-slicing, multi-process service (MPS)\redcite{NVIDIA-MPS}, and multi-instance GPU (MIG)\redcite{li2022miso} are supported by NVIDIA, with only MIG ensuring complete performance and memory isolation. MIG divides SMs into seven and memory into eight slices, which can be combined into configurable GPU instances or partitions. The configurations, referred to as {\em MIG profiles}, permissible for the MIG partitions in an A100 with 80 GB memory are provided in Table~\ref{tab:MIG-profiles}. As discussed in Sec.~\ref{sec:intro}, in MIG mode, GPU utilization metrics are reported at the partition level, while power consumption is available only for the full GPU. The rest of the paper discusses methods for distributing the aggregate power among the partitions.

\section{GPU Power-Performance Characterization Studies and Modeling} \label{sec:detailed}
\noindent
In this section, we examine the various factors that affect power consumption in GPUs using NVIDIA A100 and V100 systems. We use different workloads such as various CUDA matrix multiplication kernels (MATMUL)\redcite{matmultkernel}, a TENSOR workload based on GPU Burn benchmark (GPUBurn)\redcite{gpu-burn}, and Large Language Model (LLM) inference systems\redcite{li2024llminferenceservingsurvey}.
MATMUL evaluates ten matrix multiplication kernels (Kernels 1--10) across matrix sizes from 128 to 16,384, varying in algorithm and optimization levels, ranging from na\"{i}ve multiplication in Kernel~1 to the most optimized version in Kernel~10.
These sizes capture diverse computational demands, making matrix multiplication an ideal benchmark for GPU workloads.
GPUBurn stresses Tensor cores with matrix multiplication, while LLM benchmarks assess AI inference workloads. Together, they provide a comprehensive GPU performance evaluation for scientific computing and AI applications.

\subsection{Analysis of CUDA Workloads}
\label{sec:detailed-cuda}
\begin{figure}

    \begin{subfigure}[b]{1\textwidth}
        \includegraphics[scale=0.45]{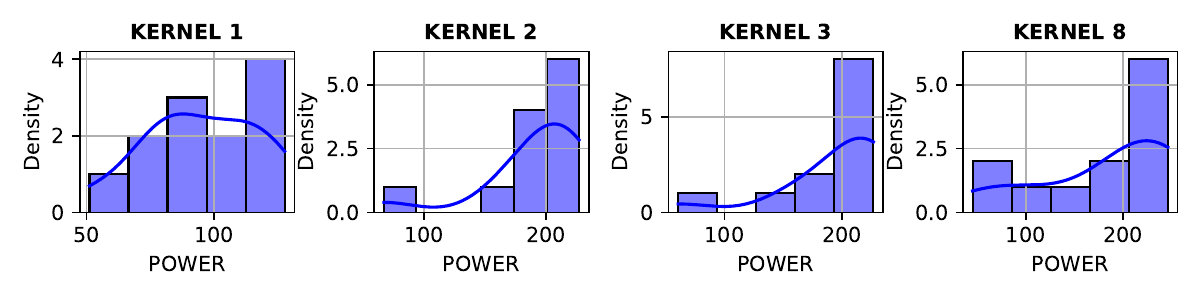}
    \end{subfigure}%

    \caption{\small Distribution of Power across kernel types}
    
    \label{fig:kerneldist}
\end{figure}

\begin{figure}

    \begin{subfigure}[b]{1\textwidth}
        \includegraphics[scale=0.45]{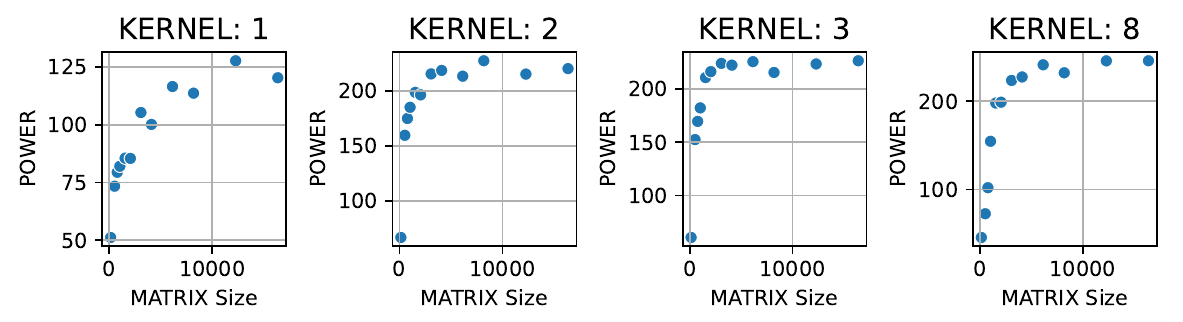}
        
    \end{subfigure}%

    \caption{\small Distribution of Power across matrix sizes}
    \label{fig:matrixsizedist}
\end{figure}

We first analyze the impact to GPU power\footnote{Power refers to instantaneous power unless otherwise specified.} due to CUDA MATMUL kernels operating on 32-bit FP matrices.\footnote{Results with other data types were mostly similar.} In this paper our focus is on determining  power for MIG instances, hence we do not consider aggregate energy. It should be noted that 
there is no implication on energy efficiency based on power consumption, since there is the possibility of a workload with lower power taking longer time to complete a given quantum of work, thereby consuming higher overall energy.

\paragraph*{\hspace{-0.78 cm} Power density analysis} Fig.~\ref{fig:kerneldist} 
plots  the power distribution for different kernels on V100 by setting clock frequency to its maximum value 1380 MHz.
The power consumption of Kernel~1 is concentrated between 50 and 125 W, with the density peaking at $\sim$125 W. For Kernel~2, with power ranging from 100 to 200 W, the density peaks significantly at $\sim$150 W, indicating higher power consumption than Kernel~1. The sharp peak suggests that the power usage is more consistent during this kernel's operation. Kernel~3's power range of 100 to 200 units is similar to that of Kernel~2 barring a broader peak.
Kernels~4--7 exhibit similar behavior to Kernel~3 and are hence omitted.
Kernels~8--10 also share a 100–200 W distribution, with Kernel~8 only depicted for brevity. The density peak, however, is slightly lower and broader, suggesting a more distributed power consumption pattern. The subtle variations and the differences in the peaks and spreads of the density curves of these three kernels reflect variations in kernel operations on different matrix sizes.
Fig.~\ref{fig:matrixsizedist} examines power by matrix size with power initially rising with matrix size before saturating, likely due to compute or memory bottlenecks or GPU power limits, which trigger automatic SM frequency scaling.

\begin{figure}

    \begin{subfigure}[b]{1\textwidth}
        \includegraphics[scale=0.3]{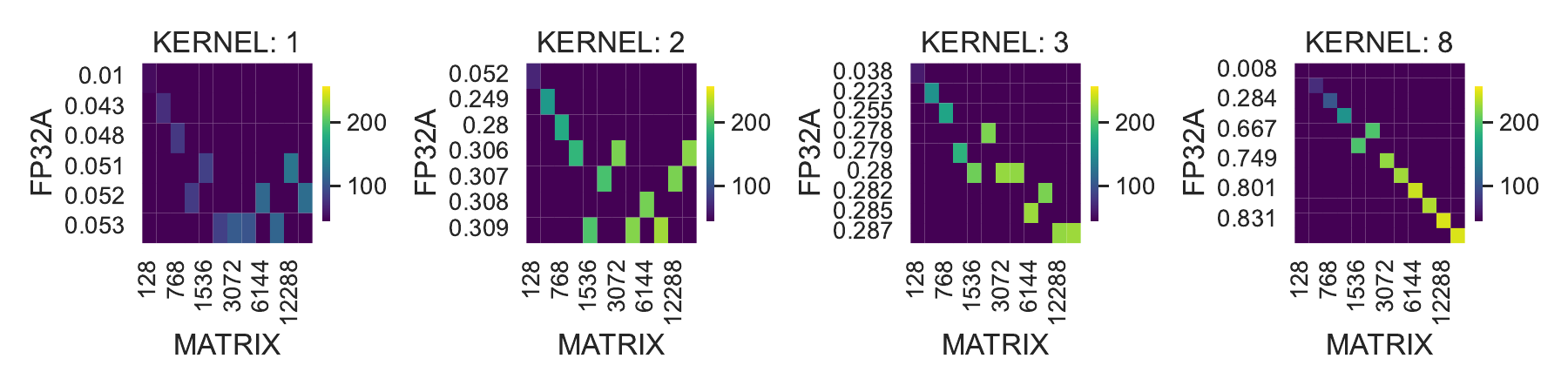}
    \end{subfigure}%

    \caption{\small Power distribution for FP32A / Matrix Sizes}
    
    \label{fig:powerdist1}
\end{figure}

\begin{figure}

    \begin{subfigure}[b]{1\textwidth}
        \includegraphics[scale=0.3]{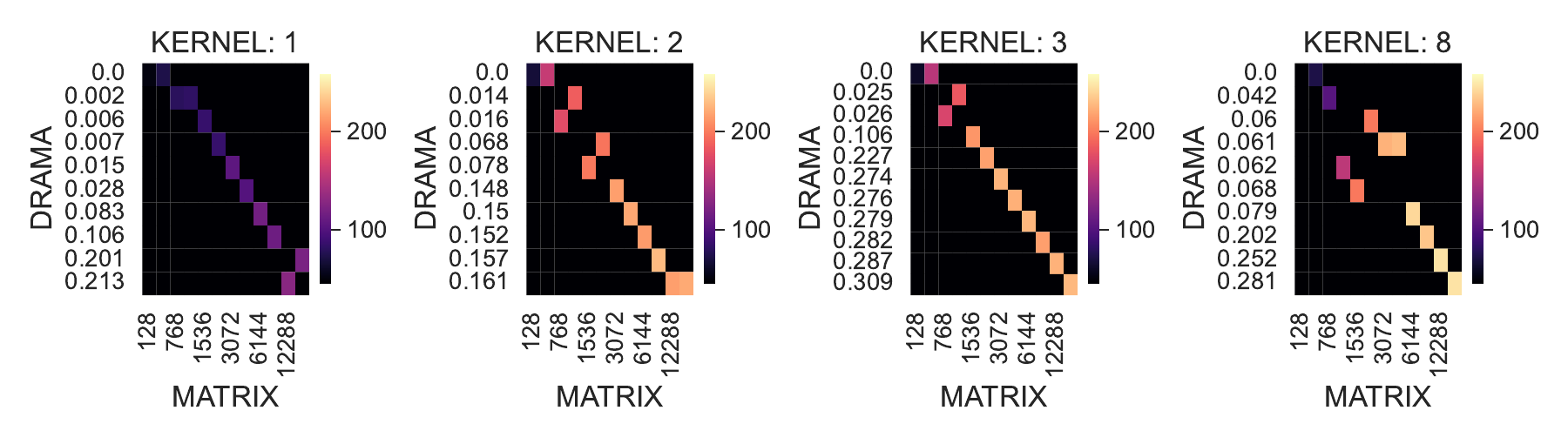}
    \end{subfigure}%

    \caption{\small Power distribution for DRAMA / Matrix Sizes}
    
    \label{fig:powerdist3}
\end{figure}

\paragraph*{\hspace{-0.75cm} Utilization Analysis} 
We next analyze GPU utilization metrics to account for the differences in the power consumption of kernels.
A correlation analysis of about 15 metrics with POWER revealed significant correlation for SMACT, FP32A, and DRAMA. We had locked the GPU clock to 1380 MHz, and hence, there was no correlation between POWER and SMCLK. Since SMACT is a coarse metric that tracks the fraction of time at least one core of any SM is active, it does not provide much insights and hence has been omitted in the analysis of this section.
The FP32A heatmaps of Fig.~\ref{fig:powerdist1} indicate the utilization ranges for FP32 CUDA cores for the various kernels, while Fig.~\ref{fig:powerdist3} depict the ranges for DRAM usage. 
FP32A values are higher for larger matrices since they can utilize a larger fraction of SM cores concurrently. It can also be seen that power consumption increases with FP32A.

The DRAMA heatmaps show that DRAM usage, similar to FP32 activity, increases with matrix size, as larger matrices occupy a larger amount of memory and also lead to larger transfers of data between the memory and on-chip caches.
In conjunction with the increase in DRAM activity, there is a noticeable rise in power consumption as matrix sizes grow. The heatmaps highlight that kernels with higher DRAM usage also tend to exhibit higher power consumption, indicating a strong correlation between memory activity and energy demands. DRAMA coupled with FP32A highlights the critical role of both memory and processing power in determining overall power consumption.

\begin{figure}[t]

    \begin{subfigure}[b]{1\textwidth}
        \includegraphics[scale=0.51]{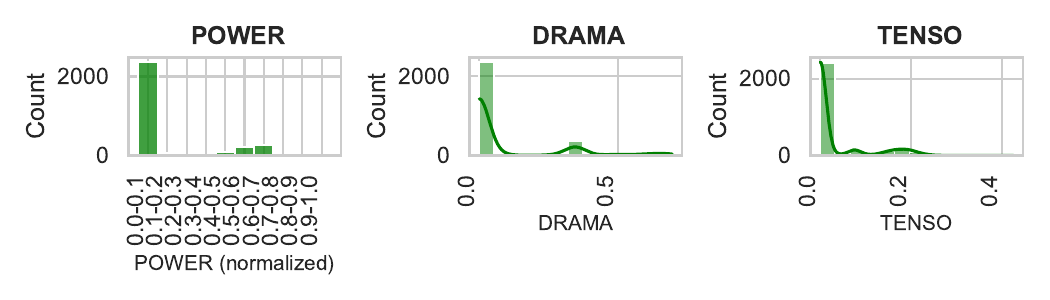}
        \vspace{-0.4cm}
        \caption{\small LLAMA LLM Workload}
    \end{subfigure}

    \begin{subfigure}[b]{1\textwidth}
        \includegraphics[scale=0.51]{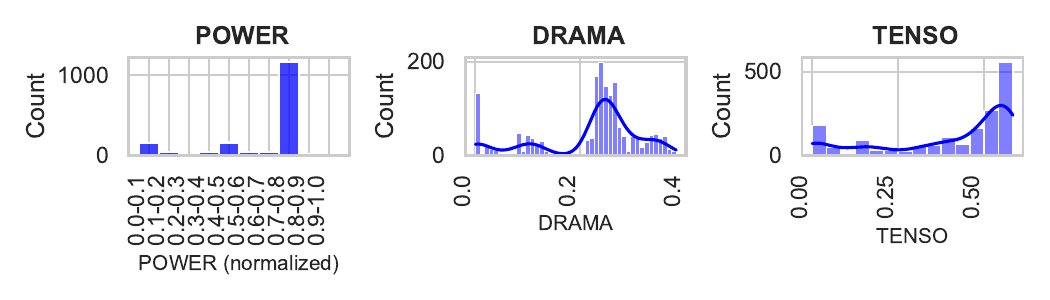}
        \caption{\small GPUBurn Benchmark Workload}
        \vspace{-0.4cm}
        \label{fig:ns65}
    \end{subfigure}%
    \caption{\small Distribution of Metrics across software workloads}
    \vspace{-0.0cm}
    \label{fig:distributionsoftwareworkloads}
\end{figure}


        

    

\begin{wrapfigure}{r}{0.28\textwidth}
        \centering
        \includegraphics[scale=0.27, trim={0cm 1.5cm 1.0cm 2.5cm},clip]{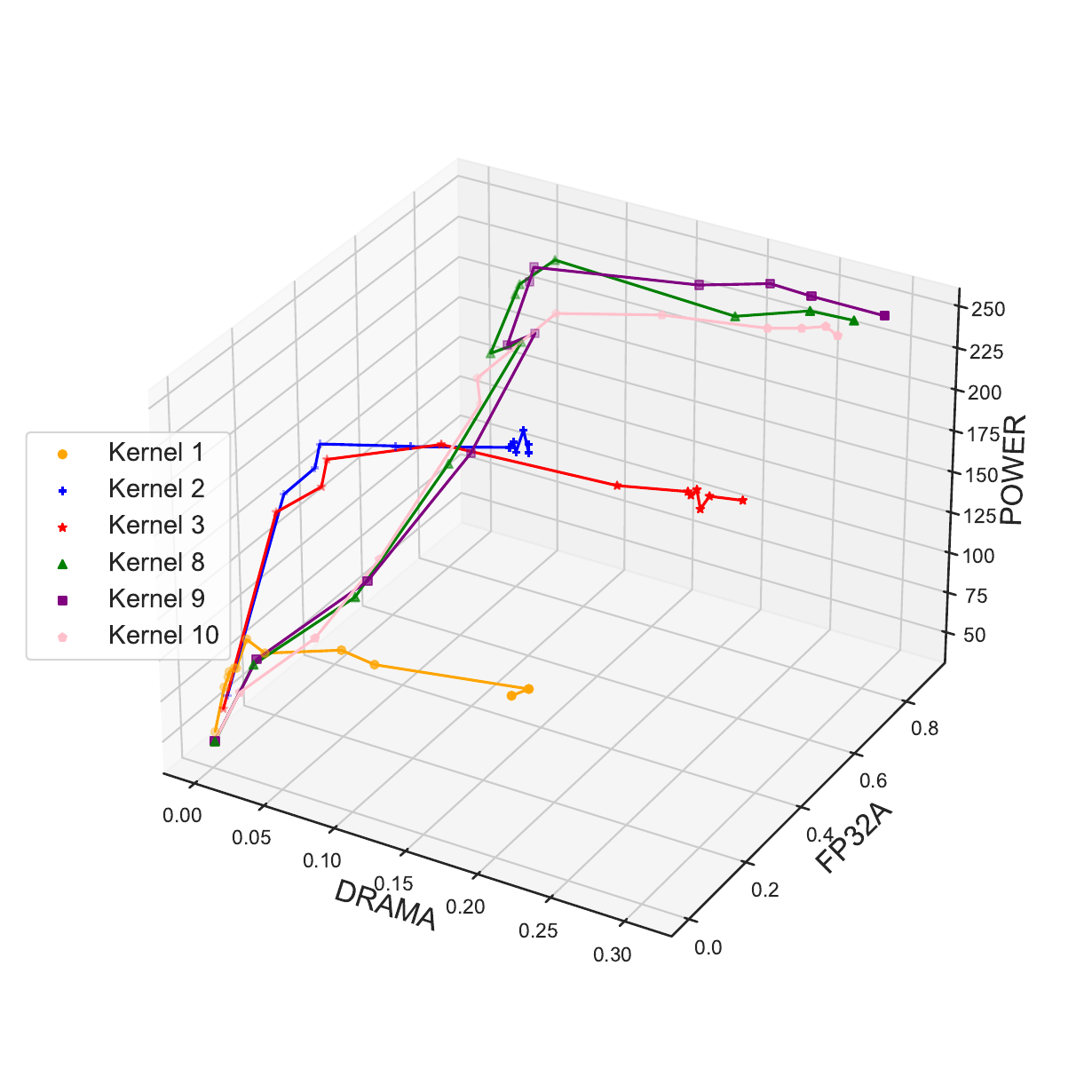}   
    \caption{\small GPU Power vs. FP32A and DRAMA for MATMUL kernels}
    \vspace{-0.6cm}
    \label{fig:3dplot2new}
\end{wrapfigure}
\paragraph*{\hspace{-0.75cm} {Utilization vs. Power Characteristics}} The 3D plot in Fig.~\ref{fig:3dplot2new} illustrates the interplay between 
DRAMA, FP32A, and POWER for different GPU kernels. For Kernel~1, a gradual increase in power consumption is observed as both FP32A and DRAMA rise. The values for FP32A are confined to the range of 0.01 to 0.05, with POWER spanning from 60 to 120 W.
For Kernel~2, the power levels are noticeably higher, ranging from 100 to 200 W. DRAMA values remain below 0.15, indicating that this metric contributes less significantly to power variation. Kernel 3 shows a similar trend.
For Kernels 8--10, with a broader range for both FP32A (0.0 to 0.8) and DRAMA (0.0 to 0.3), power  varies significantly, reaching up to 200 W.
Across all kernels, power saturates after peaking with FP32A, with further increases in DRAMA having no impact. Power scales linearly with FP32A but with varying slopes—highest for Kernels~1--3, which are the least optimized. Note also that there does not seem to be any relation between DRAMA and the slope, since  the slope of Kernel~3 with lower DRAMA than Kernels~8--10 is higher, but is lower than Kernel~1 whose DRAMA is lower than itself.

\begin{wrapfigure}{r}{0.28\textwidth}
        \includegraphics[scale=0.28]{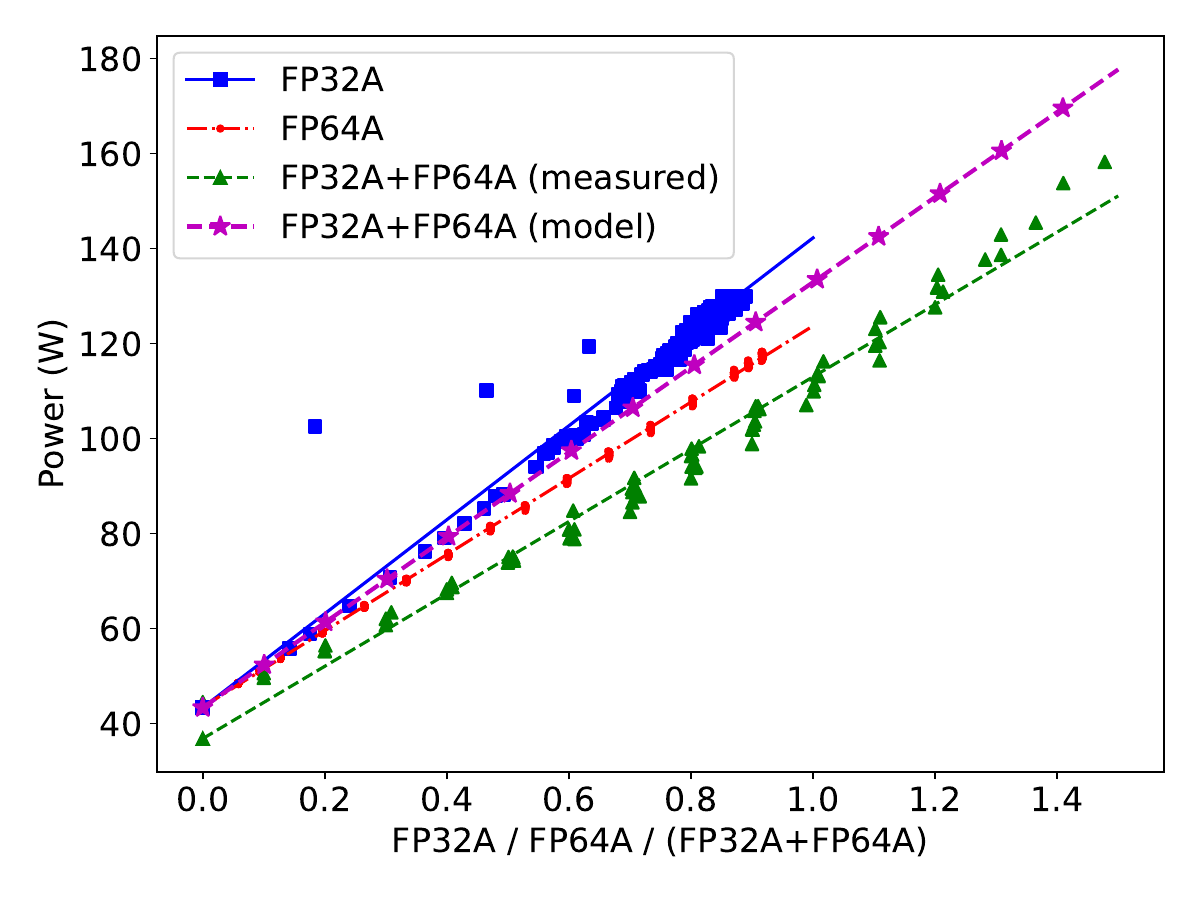}   
    \caption{\small GPU Power for independent and concurrent execution of FP32 and FP64 vector operations}
    \label{fig:FP32-FP64-additivity}
    \vspace{-0.15cm}
\end{wrapfigure}

\paragraph*{\hspace{-0.6cm}Concurrent Kernel Execution}We now analyze the power impact of concurrently executing FP32 and FP64 vector kernels while minimizing memory influence by keeping data entirely on-chip. By independently exercising FP32 and FP64 cores and varying thread dispatch, we control utilization to assess their individual and combined power consumption effects. Power consumption by core utilization for these independent executions  are shown by the blue and red plots, resp., in Fig.~\ref{fig:FP32-FP64-additivity}. 
The relationship is linear for both due to the absence of DRAM accesses. We next executed the two workloads concurrently, increasing their utilizations systematically. The utilization-power relation for the combined workload is also shown in the same figure (by the green plot).
(The total utilization exceeds 1.0, since the utilization for each core type can be up to 1.0.) 

Note that the combined power for a given total utilization is less than the power consumed by each workload at that utilization level, which by itself suggests that the power consumption of the two workloads may not be additive. For validation,  we used linear regression models constructed for the individual workloads (blue and red plots) and used them to compute the expected power for their combined execution. The purple plot shows this expected power, which is much higher than the actual power consumed, further corroborating that FP64 and FP32 powers are not additive. However, this behavior was not so pronounced when both workloads used either FP32 or FP64 cores only. 
Thus, additivity of workloads cannot be expected to hold at all times, which 
argues against the the possibility of simple and lightweight power partitioning approaches for MIG.

\subsection{Analysis of Tensor Workloads}
\label{sec:detailed-tensor}
\noindent
Having demonstrated the heterogeneity even within CUDA kernels accomplishing the same end task, we now analyze TENSOR kernels.
Fig.~\ref{fig:distributionsoftwareworkloads} shows the distribution of key metrics of POWER, TENSO, and DRAMA across the two workloads of LLM Inference with Llama-3B model (LLAMA) and GPUBurn performing matrix multiplications, both  on A100. 
The power consumption during GPUBurn is higher, with a broader distribution, reflecting the benchmark workload's impact on power. Similarly, DRAMA is also higher for GPUBurn on average as opposed to LLAMA, however comprising mild peaks at $\sim$0.4. The TENSO activity on LLAMA is also lower on average as compared to GPUBurn. These differences between the two sets of distributions highlight the lack of generalization  across workloads with each workload having its own set of characteristics. This imbalance renders it more difficult to accurately model the system's behavior across all possible workload conditions. Consequently, the variability renders having a universal model towards power prediction a significant predicament, particularly with the black-box visibility into workloads in execution.


\begin{figure*}

    \begin{subfigure}[b]{1\textwidth}
        \centering
        \includegraphics[scale=0.4]{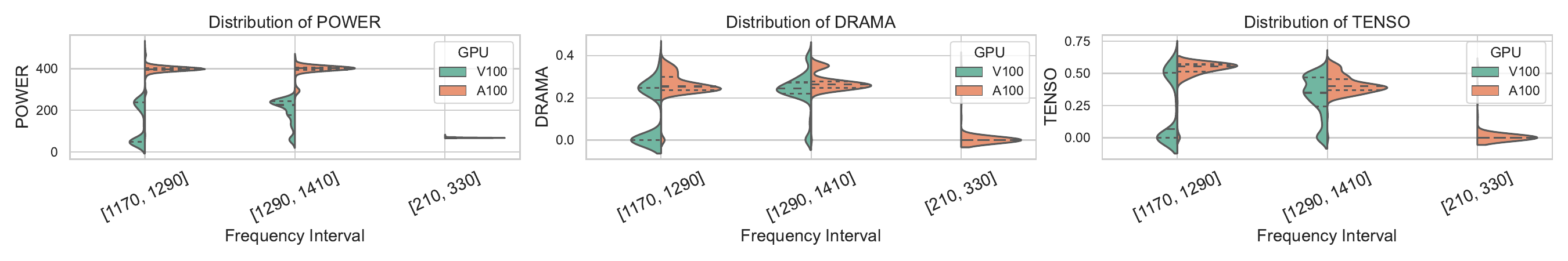}
    \end{subfigure}%
    
    \caption{\small Hardware heterogeneity on metrics}
    
    \label{fig:a100violin}
\end{figure*}

\subsection{Hardware Heterogeneity}
\label{sec:detailed-hw-het}
\noindent
In order to demonstrate the impact of s/w-h/w interplay, we execute GPUBurn on both A100 (with 80~GB RAM) and V100 (with 40~GB RAM), leaving the clock frequency unlocked. Fig.~\ref{fig:a100violin} provides a comparative analysis between the two.
A100 consistently operates at higher power levels compared to V100 due to its higher computational capabilities (such as larger number of SM's) as well as larger memory. 
In terms of DRAMA, both A100 and V100 show similar distributions as the same  workload is used on the two GPUs, indicating the software stack's influence on the metric. A100 shows significantly higher TENSO values at higher frequencies compared to V100, highlighting its architectural enhancements for tensor operations (e.g., support for sparsity, larger tensor cores). 
Interestingly, TENSO values exhibit peak activity at the highest frequency range of operations as opposed to the mid  frequency ranges depicting better utilization of tensor cores at high frequencies.

\begin{figure}

    \begin{subfigure}[b]{1\textwidth}
        \includegraphics[scale=0.35]{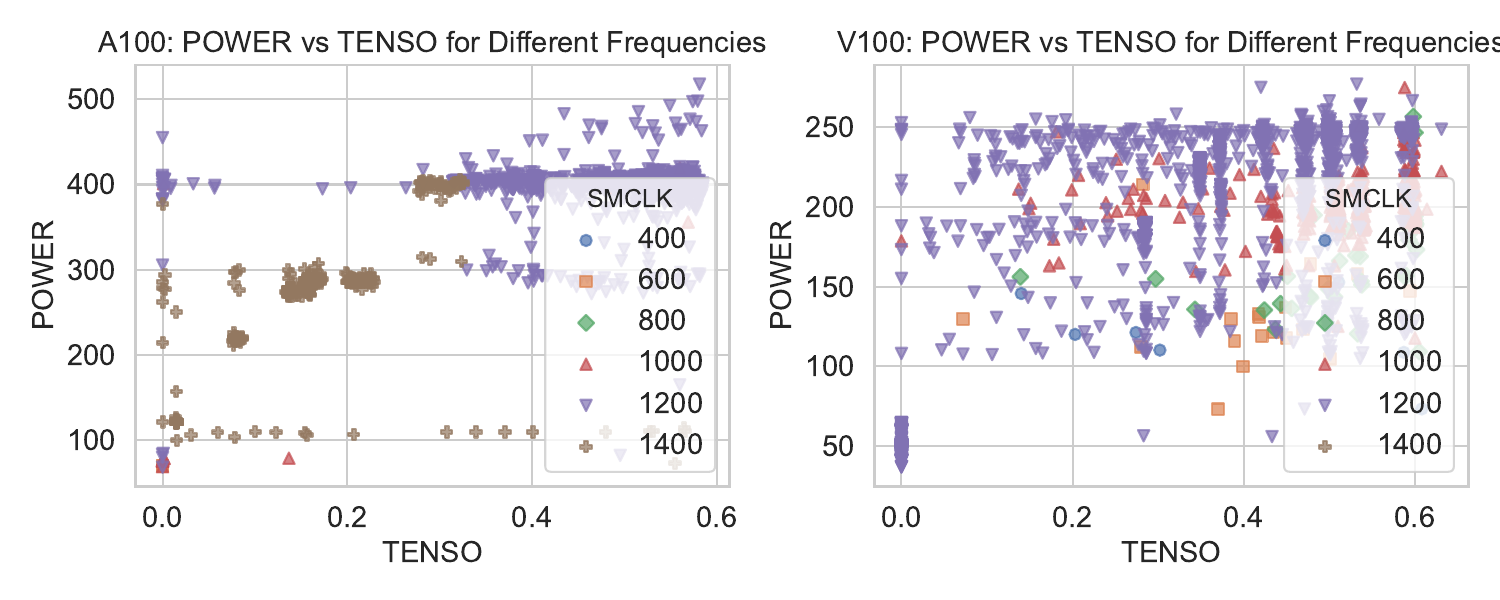}
    \end{subfigure}%

    \caption{\small Distribution of metrics across hardware}
    
    \label{fig:distributionhardware}
\end{figure}

Fig.~\ref{fig:distributionhardware} shows hardware-specific impacts on the same workload, focusing on tensor activity (TENSO) during GPU Burn. On  A100, power increases with TENSO, reaching up to 500 W at higher frequencies (e.g., 1200–1400 MHz), while lower frequencies limit power to 100–200 W. Dense clusters reflect consistent relationships between computational intensity and power, though idling periods at low TENSO values result in lower power at high frequencies.
In contrast, V100 shows lower power levels overall, maxing at 250 W for frequencies of 1200–1400 MHz, with narrower data spreads. These differences highlight A100's advanced architecture handling higher power loads versus V100's constrained power profile.
%

Dynamic variability of runtime monitored metrics on heterogeneous GPU hardware stems from disparities in monitoring granularity and resource allocation visibility across instances. Some workloads show distinct metric patterns due to high resource utilization, while others are underrepresented, leading to uneven data distributions. This imbalance complicates predictive modeling by failing to capture nuanced workload-GPU interactions. Ensuring consistent power prediction thus necessitates tailored approaches to handle workload heterogeneity dynamically at runtime.

\begin{figure}

    \begin{subfigure}[b]{1\textwidth}
        \includegraphics[scale=0.43]{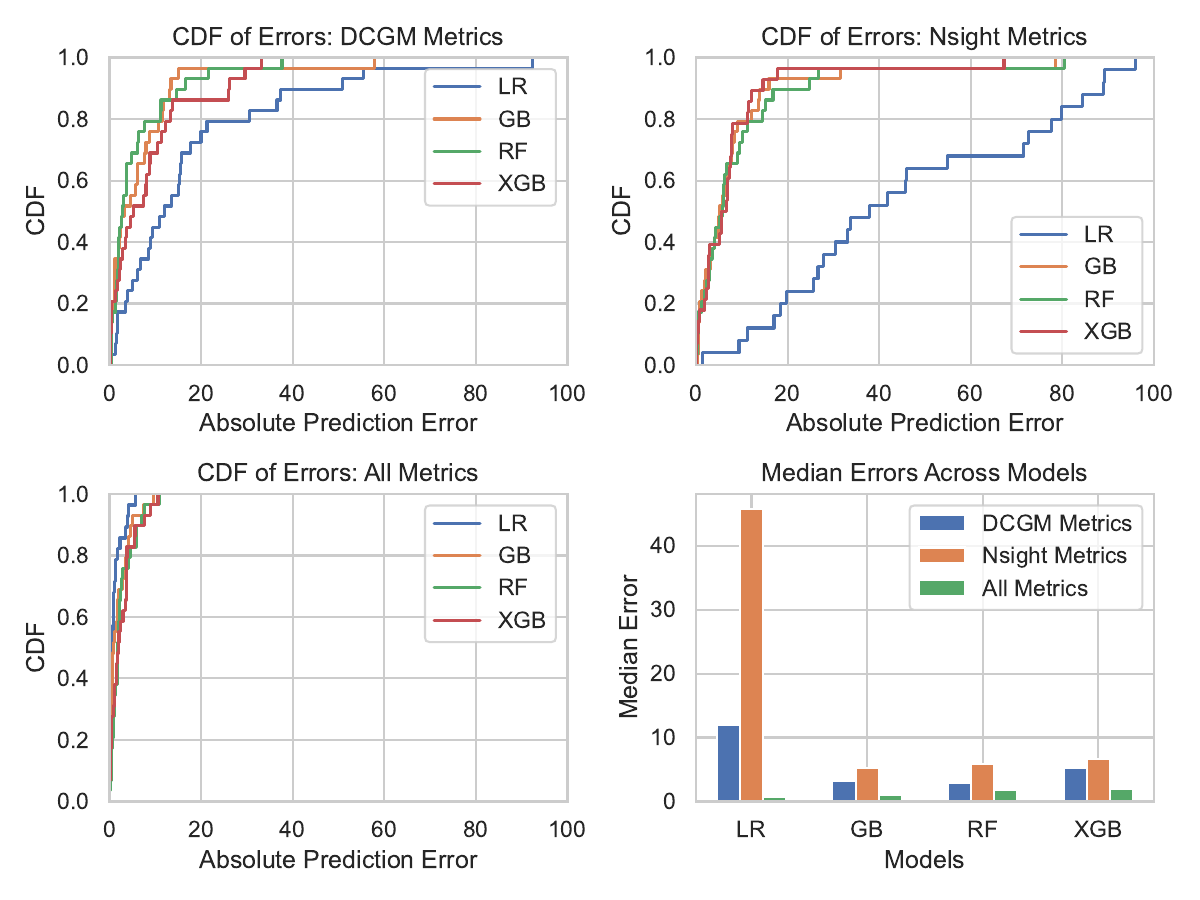}
    \end{subfigure}%
    
    \caption{\small DCGM vs. NSIGHT compute metrics}
    \vspace{-0.2cm}
    \label{fig:dcgmnsight}
    \vspace{-0.2cm}
\end{figure}

\subsection{Implications of Metrics}
\label{sec:detailed-metrics}

\noindent
While DCGM metrics are lightweight and easily available, NVIDIA Nsight Compute (NCU) metrics provide more advanced statistical measures into the working of the GPU components. However, obtaining such metrics involve large overheads that makes it impossible to aquire them in production systems. In order to quantify the benefits provided by NCU metrics on the power consumption specifically, we train four models, Linear Regression (LR), Gradient Boosting (GB), Random Forest  and XGBoost on the MATMUL workload. 
We chose the NCU metrics with non-trivial correlation with power for inclusion in the models. DRAM and cache access related metrics showed the highest correlation.
Fig.~\ref{fig:dcgmnsight} demonstrates the benefit of incorporating NCU metrics versus only relying on DCGM metrics. When using only NCU metrics, there is no significant improvement, indicating that these metrics alone may not provide substantial additional predictive power. The combined graph integrating both DCGM and NCU metrics, however shows significant improvement, suggesting that the inclusion of NCU metrics offers benefits beyond what is achieved with DCGM metrics alone, producing much lower median error. However, NCU metrics add significant overhead and are difficult to get in production systems. Furthermore, more studies are required to ascertain their value in more complex workloads using tensor cores. So we leverage lightweight DCGM metrics towards power attribution in partitioned GPUs.

\subsection{Full GPU Power Models}
\label{sec:detailed-training}


\begin{figure*}

    \begin{subfigure}[b]{1\textwidth}
        \includegraphics[scale=0.4]{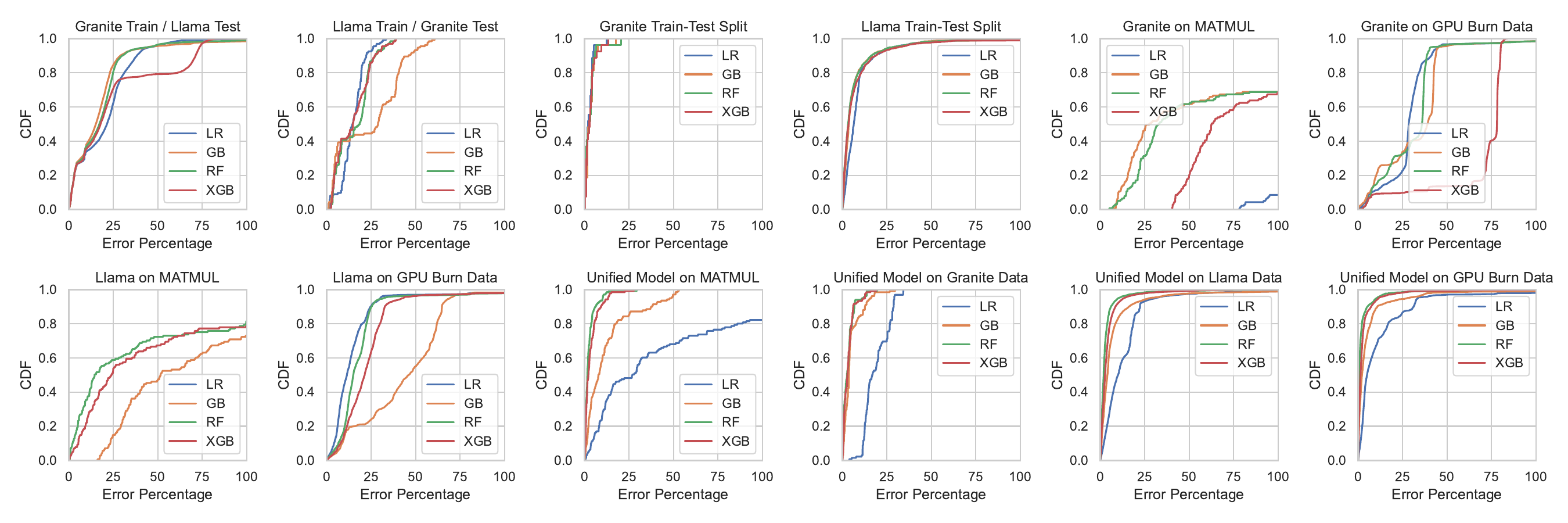}
    \end{subfigure}
    
    \caption{\small CDF of errors for model prediction}
    
    \label{fig:cdferrors}
\end{figure*}



    
    
\small
\begin{wraptable}{l}{0.28\textwidth}
\caption{\small Training Time Summary (7435 sample size)}
\label{tab:overhead}
\scalebox{0.9}{
\begin{tabular}{lc}
\toprule  
Model & Training Time (s) \\
\midrule
Linear Regression & 0.0017 \\
Gradient Boosting & 0.5671 \\
Random Forest & 1.7844 \\
XGBoost & 0.0713 \\
\bottomrule
\end{tabular}
}
\end{wraptable}
\normalsize
\noindent
Fig.~\ref{fig:cdferrors} presents CDF (Cumulative Distribution Function) plots showing error percentages for models trained and tested across different datasets. We evaluate and compare LR, GB, RF, and XGB models. For the Granite Train/Llama Test plot, all four models demonstrate similar performance, reflecting their ability to generalize across workloads with shared characteristics.
When training and testing on the same dataset (e.g., Granite Train/Granite Test or Llama Train/Llama Test), errors are low, highlighting the strong fit of the models to consistent data sources. These scenarios reveal minimal variability, demonstrating how well the models adapt to familiar patterns. In contrast, cross-validation across heterogeneous workloads, such as Granite Train/MATMUL Test, results in higher prediction errors. The largest errors occur in the Llama Train/MATMUL Test scenario, reflecting the challenges models face when predicting power consumption for workloads vastly different from the training data. LR struggles the most in these cases, with error percentages often exceeding $\sim$90\%, while GB and RF fare somewhat better but still exhibit variability.
Cross-validation tests further underline the importance of training data diversity. Models trained on granite and tested on GPUBurn or MATMUL show a clear increase in error percentages, particularly for LR and GB. Table~\ref{tab:overhead} demonstrates the overhead associated with training the different types of models. The overhead runtime depicts the low computational overheads and shows that the methodology can be adapted in an online setting.

To improve generalization, mitigate errors and enhances the ability of models to handle diverse scenarios effectively, it is crucial to curate representative training sets that encompass a wide range of workload behaviors. 
The variability across workloads and predictor models complicates creating a unified offline model for diverse workloads. A unified model trained on combined datasets (Granite, Llama, MATMUL, GPUburn) captures both hardware behavior and workload-specific characteristics, resulting in the lowest error. This shows that aggregated data enables better prediction than isolated datasets. While LR struggles to capture non-linear GPU metric-power relationships, RF proves to be the most robust predictor, highlighting that no single methodology fits all for power prediction across heterogeneous workloads.

\section{Partitioning GPU Power}
\label{sec:GPU-power-partitioning-MIG}
\noindent
In this section, using GPU power characteristics learned through benchmarking, we design approaches for partitioning the aggregate GPU power among its MIG partitions. These approaches can be adapted with minimal changes to other forms of shared execution such as time-sharing and {\em multi-process service\/} (MPS) also. The high-level principles should be applicable in general to GPUs from other manufacturers such as Intel and AMD.

As already discussed in Sec.~\ref{sec:detailed-cuda} (see discussion around Fig.~\ref{fig:3dplot2new}), a simple, static-weight based linear estimation of GPU power or  partitioning it in proportion to the utilization of the various sub-components would not have good accuracy.  Hence, we apply increasingly complex methods, which are presented and evaluated below. 
First, we define {\em active\/} and {\em idle power\/} and explain how these are handled during power partitioning. 
\paragraph*{\hspace{-0.75cm}Active and idle power} Compute processors, including GPUs, are known to suffer from lack of {\em energy proportionality\/}\redcite{energy-prop}, and consume power even in idle state when no process is active. Such power is referred to as {\em idle power\/}, which is generally a non-trivial amount. The idle power of  NVIDIA A100-80GB 
is $\sim$85 W at frequencies above 1200 MHz against its peak power of $\sim$400-450 W.
When a GPU executes some process, additional power is consumed, which is referred to as the GPU's {\em active power\/}. This active power is dependent on the extent to which different units of the GPU are exercised. A GPU's {\em total power\/} is  the sum of its idle and active powers. 

In a GPU with MIG partitions, the idle power is common to all the partitions, and since it is independent of process execution, its distribution among the partitions should  be independent of the MIG utilization levels. Hence, we divide the idle power in proportion to the sizes of the partitions. Thus, a MIG instance with $k$ compute slices, denoted as a $k$G partition, where $k$ is one of 1, 2, 3, 4, or 7,  will be allocated $\frac{k}{n}^{\rm th}$ of idle power, with $n$ being the total number of slices in the partitions with job assignments.
 The idle power of a GPU can be noted for different clock frequencies or it can be estimated by substituting all utilization parameters to 0.0 in the GPU's power model. The rest of the section deals primarily with active power distribution.

Another MIG-specific aspect is that utilization metrics reported by DCGM for a MIG partition are with respect to the MIG's capacity. Hence, the metrics need to be normalized with respect to the full GPU for estimating their power using models. Thus, the metrics of a $k$G instance are normalized using $\frac{k}{n}$ as the scaling factor, where $n$ is the total size of all the partitions (not just the ones with job assignments).

We are now ready to present various methods for determining MIG partition power.

\subsection{Estimation via Generic Full GPU Power Models}
\label{sec:generic-model-part}
\noindent
This method uses the unified model (which is constructed using training data from multiple representative workloads) as described  in Sec.~\ref{sec:detailed-training} to estimate  MIG instance power. First, the unified model is used to predict the power consumption of each MIG instance individually, with the instance's normalized DCGM metrics serving as the model features. Thus, each MIG prediction is for the power that would be consumed when just that MIG's workload executes on the GPU, and hence would include the full GPU idle power. Hence,  to obtain the MIG's active power, full GPU idle power is deducted from the total predicted power. Next, idle power for each MIG instance is computed in proportion to its size and the total size of all the active instances as described at the beginning of the section.
For example, suppose that only a 2G and 3G instances are active, then they will be assigned $\frac{2}{5}^{\rm th}$ and $\frac{3}{5}^{\rm th}$ of the idle power, resp. 
Finally, the total power for an instance is obtained by adding its idle and active powers.
The accuracy of the overall approach is evaluated by summing all instance powers and comparing the sum to the total measured GPU power. 

\begin{figure}[htp]

    \begin{subfigure}[b]{1\textwidth}
        \centering
        \includegraphics[scale=0.3]{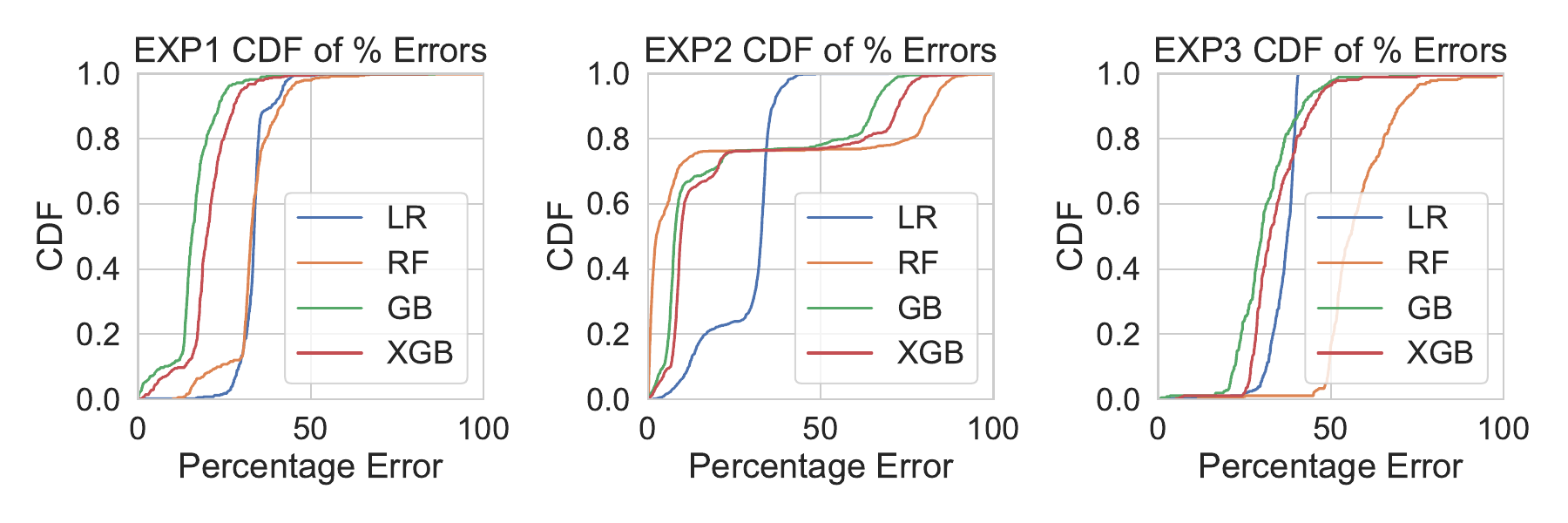}
    \end{subfigure}
    
    
    
    \caption{\small CDFs for MIG power predictions with unified training data on non-identical workloads}
    \label{fig:cdfmig}
    \vspace{-0.4cm}
\end{figure}

\begin{wraptable}{l}{0.3\textwidth}
    \centering
    \scalebox{0.8}{
    \begin{tabular}{|c|c|c|}
    \hline
    \textbf{EXP} & \textbf{MIG-2G} & \textbf{MIG-3G} \\ 
                 & \textbf{Partition} & \textbf{Partition} \\ 
    \hline
    EXP1 & BURN 2G & LLAMA 3G \\ \hline
    EXP2 & FLAN 2G & GRANITE 3G \\ \hline
    EXP3 & GPUBURN 2G & GPUBURN 3G \\ \hline
    \end{tabular}
    }
    \caption{\small MIG Instance combinations in MIG paritions}
    \vspace{-0.2cm}
    \label{tab:exp_descriptions}
\end{wraptable}
Tab.~\ref{tab:exp_descriptions} shows the various combinations of workloads used to evaluate the approach. Fig.~\ref{fig:cdfmig} presents the CDFs of error \%'s for various ML models (trained using the unified dataset) predicting power for different MIG combinations. To test the model's generalizability, we evaluated workloads with a different training set from those used in Tab.~\ref{tab:exp_descriptions}, namely BLOOM and MATMUL benchmark.  As in earlier scenarios, model performance varies with workload characteristics. Notably, in MIG-specific tests, EXP2 has the lowest median comparative error amongst the three experiments. We observed that the range of metrics for EXP2 spans a very low range towards the lower end of the spectrum,  $\sim$95\% of the data points. This reflects the ability of non-identical workloads to generalize better towards fragmented ranges as opposed to generalizing well across the broad metric range spectrum.

\begin{figure}[htp]

    \begin{subfigure}[b]{1\textwidth}
        \centering
        \includegraphics[scale=0.3, clip]{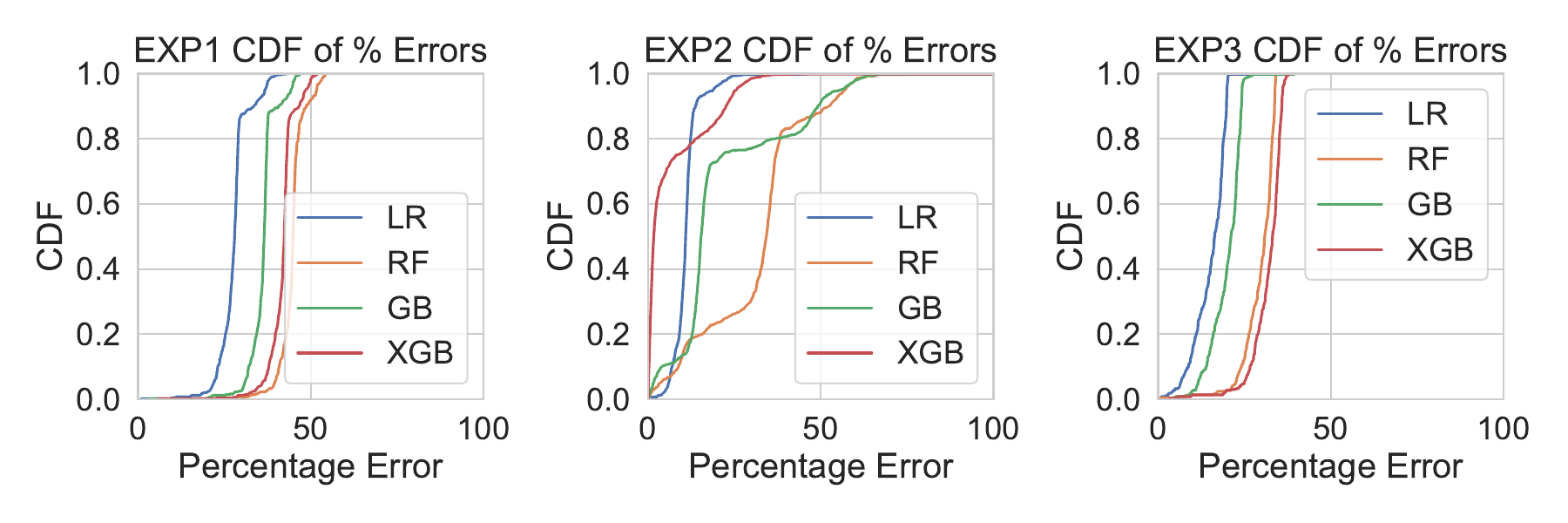}
    \end{subfigure}
    
    
    
    \caption{\small CDFs for MIG Power predictions with Unified training data on identical wokloads}
    \label{fig:cdfmig2}

\end{figure}

In Fig.~\ref{fig:cdfmig2}, the CDFs illustrate prediction errors when the unified model is trained  on execution data collected from the full GPU running the same workloads. These results highlight that using workloads executed directly on the MIG instances as predictors improves the accuracy of power predictions on average across the experiments. Across the experiments, each prediction model varies in accordance with the different workloads with LR performing best in EXP3 whilst XGB and RF exhibit similar behaviour.

\subsection{Estimation via Workload-Specific Full GPU Power Models}
\label{sec:workload-model-part}
\noindent

\begin{figure}[htp]

    \begin{subfigure}[b]{1\textwidth}
        \centering
        \includegraphics[scale=0.3]{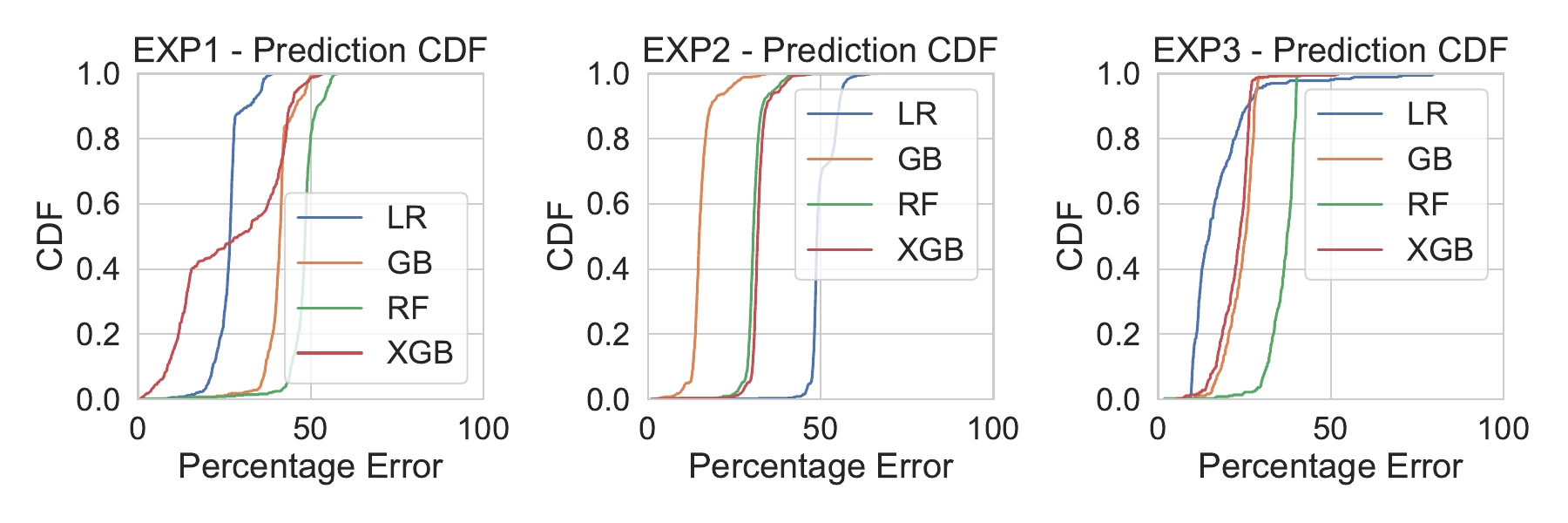}
    \end{subfigure}
    
    
    
    \caption{\small CDFs for MIG Power predictions with specific workload data as trained model}
    \label{fig:cdfmig3}

\end{figure}

Fig.~\ref{fig:cdfmig3} showcases CDFs for prediction errors when models are trained using workload data specific to the MIG instances. That is,
different models are used for predicting power for different MIG instances, where training data for a MIG's model is derived from a workload that is the same as that on the MIG instance but running on a full GPU. 
Models constructed with MIG workload-specific metrics perform, on average, similarly to those trained on unified data. However, these workload-aware models incur high overhead due to the requirement of having individual workload-specific information available at runtime, which is not always feasible. Runtime workloads often present an opaque view, making it challenging to identify similar workloads. This highlights the advantages of a unified training approach.

\begin{figure}[t]
        \includegraphics[scale=0.34]{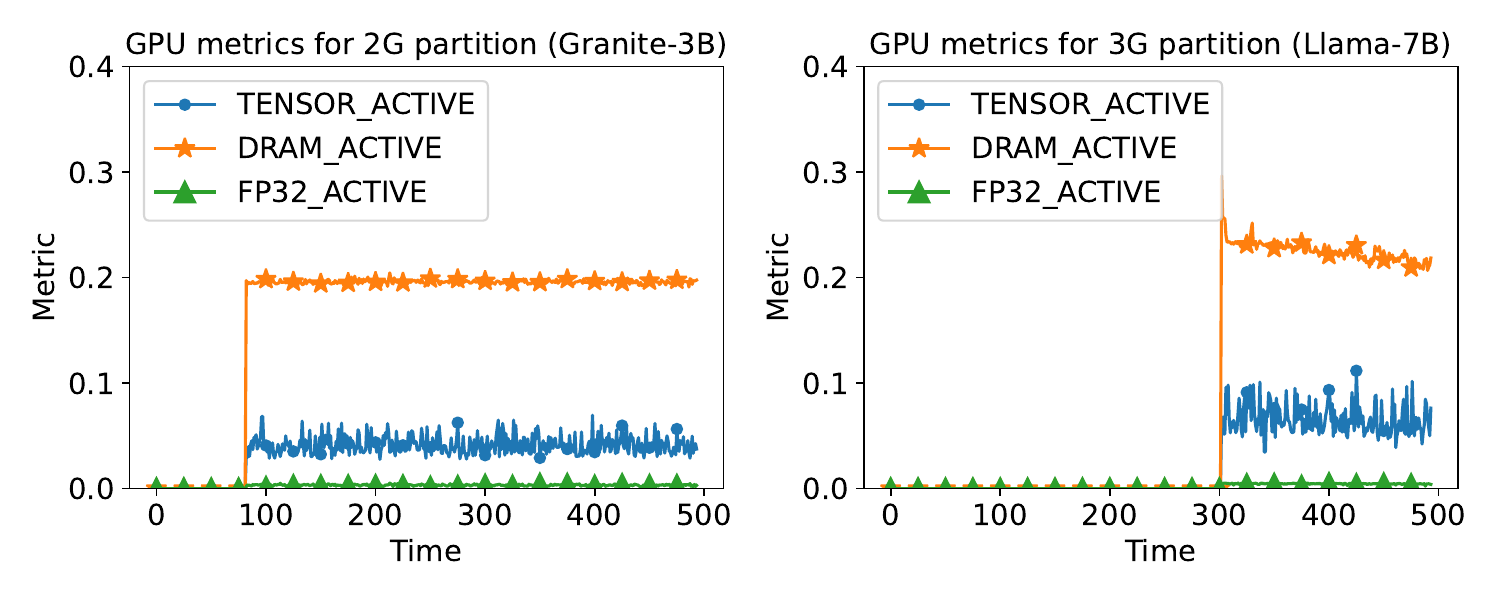}
    \caption{\small DCGM metrics for overlapping executions of Granite and Llama inferencing models on 2G and 3G partitions.} 
    \vspace{-0.5cm}
    \label{fig:MIG-2G-3G-DCGM-metrics}
\end{figure}

\subsection{Partitioning Aggregate Power Using Estimates}
\label{sec:MIG-power-scaling}
Results of the previous section show that direct estimation of power for MIG instances using GPU power models can be  error-prone. Note that the error in the case of MIG is higher than for full GPU, as errors of individual predictions add up. 

\begin{figure*}
    \hspace{-0.18cm}\includegraphics[scale=0.4]{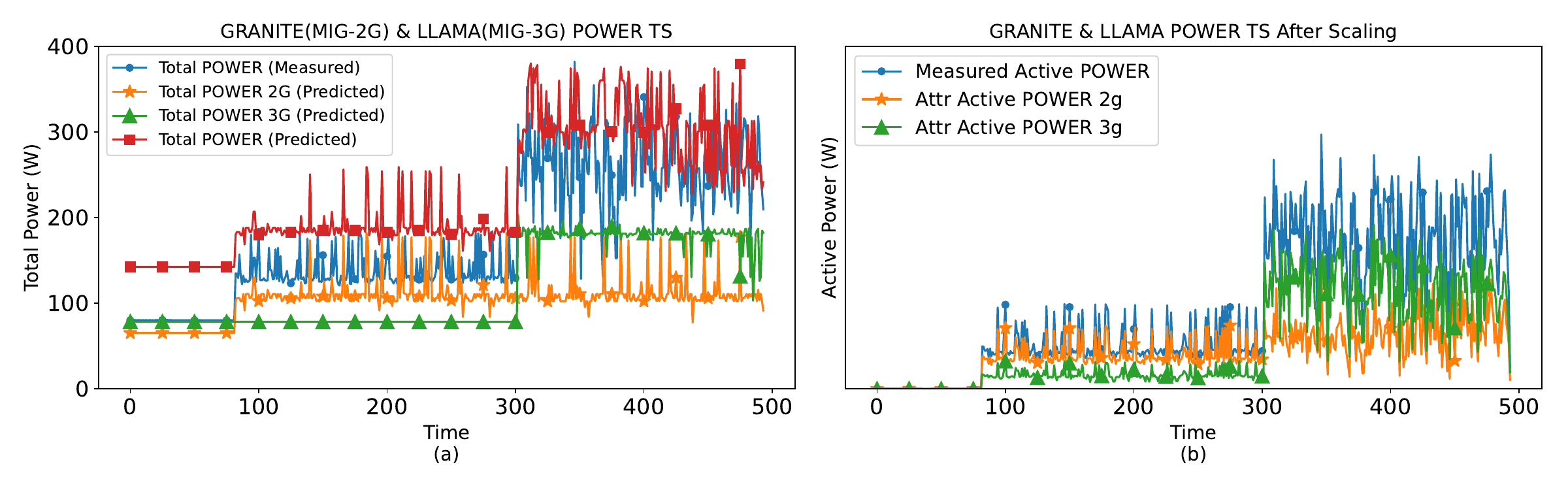}
    \caption{\small {\bf(a)} Total measured GPU power and predicted per-partition and total power for 2G and 3G partitions.
    {\bf (b)} Active power attribution to 2G and 3G after scaling.} 
    \vspace{-0.15 cm}
    \label{fig:MIG-2G-3G-power-combo}
    \vspace{-0.3cm}
\end{figure*}

We now propose a method for reducing the error in attributing power to MIG instances.
For this, we perform a controlled experiment using a combination of Granite-3B and Llama-7B inferencing workloads running concurrently 
 on a 2G and 3G partition, resp. We do not load the 3G workload 200 time units
after loading 2G. SM core and DRAM utilizations
of the workloads are shown in Fig.~\ref{fig:MIG-2G-3G-DCGM-metrics}. Power for the two MIG instances are computed as described in Sec.~\ref{sec:generic-model-part}.
 The total measured power and the power
values predicted for the two workloads using a non-identical unified model 
along with their sum (total predicted power), are shown in
Fig.~\ref{fig:MIG-2G-3G-power-combo}(a). As can be seen, the sum of the two MIG powers is significantly higher than the measured power, with the mean
average \% error (MAPE) at $\sim$40\%. 
Most importantly, note that in the
interval [0,100), both the partitions are idle, and the total
measured power (blue plot) coincides with 3G’s prediction,
while the predicted power is higher by around 70W. 
This is 
an artifact of the inaccuracy of the power models at  low
utilizations.

One easy way to reduce the aggregate error to zero is to use the measured full GPU power, when available, to scale the estimated MIG powers in proportion to their estimated values (such that the sum of their scaled values equals the measured value). More specifically, letting $P_k^{act}$ denote the active power of the $k{\rm th}$ partition, its scaled power is given by $\frac{P_k^{act}}{\sum_i P_i^{act}}\times P_{GPU}^{act}$, where $P_{GPU}^{act}$ is the GPU's measured active power. It is easy to see that the sum of all partition powers would equal the measured powered by this method. One rationale/justification for the scaling (despite errors in the predicted values $P_i^{act}$ of the MIG partitions) is the assumption that the errors in MIG predictions are in proportion to their actual powers. However, there is no ground truth to validate the fairness of such attribution. One test of fairness and accuracy is to see how close the active power  attributed to a partition is in comparison to the power it consumes when executing alone, and whether it
%
remains unaltered when load in other partitions changes. We demonstrate that this is the case.
Note that when not scaled, the power estimated for a MIG is unaffected by the activity in other partitions, since the estimation is done independently for each partition.  

The  active power attribution to the partitions after scaling is shown in Fig.~\ref{fig:MIG-2G-3G-power-combo}(b). (We do not show the total attributed  power since it would equal the measured power by our method's design.) Since the error in the aggregate power is zero, we evaluate the accuracy of partitioning by keeping one of the workloads fixed and changing the load to the other. If our attributions are accurate, then there should not be much change to the workload that remains fixed due to changes to others that are co-located.

\subsection{Estimation and Partitioning via Online MIG Power Models}
\label{sec:window-based-part}
\noindent
Our next attempt to improve the fairness of the attribution is to construct models at run-time using the utilizations of each workload running in the MIG partitions as features, and use those models for MIG power computation. As shown in Table~\ref{tab:overhead} in Sec.~\ref{sec:detailed-training}, model training times are negligible, especially given that MIG power/carbon reports can tolerate some delay, and hence, this approach is practically viable.  With this approach, instead of using the aggregate metrics for the whole GPU as features, we expand the feature set to include the metrics of each MIG separately. Thus with $n$ partitions, there would be an $n$-fold increase to the number of features. The reason for this is that, as seen already, the power characteristics of workloads can be different. Further, power consumption can depend on the data that the workloads process~\cite{ALUPower}. Hence, the differences in the workload characteristics can be made more explicit to the model by dividing each aggregate utilization metric into its MIG-level components. Although the features are MIG-level, the target variable is the whole GPU power, since that is the only available ground truth. As with previous approaches, MIG-level power is obtained in a two-step process: MIG power in the first step is obtained via the constructed model by setting the utilizations of all the  MIGs except the one under consideration to 0.0. We then deduct the full GPU idle power to obtain each MIG's active power. These  active powers are then scaled using the method of Sec.~\ref{sec:MIG-power-scaling}. 

\begin{figure}
       \hspace{-0.2cm} \includegraphics[scale=0.22]{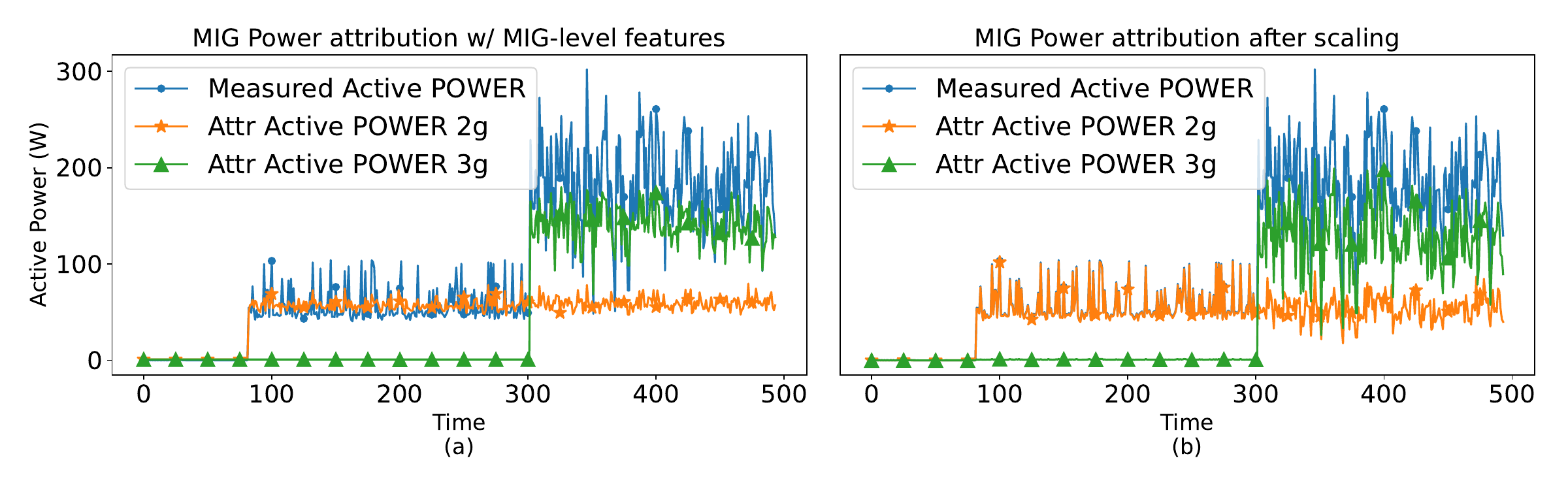}    
    \caption{\small MIG power attribution using MIG-level features.}
    \vspace{-0.2cm}
    \label{fig:MIG-2G-3G-power-window-based}
\end{figure}

Results of the method with and without the scaling step are shown in Fig.~\ref{fig:MIG-2G-3G-power-window-based}. Recall that without scaling, the total estimated active power may not equal the measured active power. This can be seen in the inset~(a), wherein during [90,100], when only 2G sees load, the total active power should equal that of 2G, but the two differ. Note that the variance in the estimated power is much lower than that in the measured power. The same holds after time 300, when 3G also starts seeing load. This issue gets resolved with scaling as seen in the inset~(b). In the segment in which only 2G has load, the active power of 2G is identical to the measured values. After time 300, the variance of 2G is slightly lowered, since the total variance is now the sum of 2G and 3G variances. However, unlike in Fig.~\ref{fig:MIG-2G-3G-power-combo}, there is no perceptible shift to the mean power of 2G at 300 when 3G load starts.

\begin{figure}
        {\hspace{-0.6cm}\includegraphics[scale=0.23]{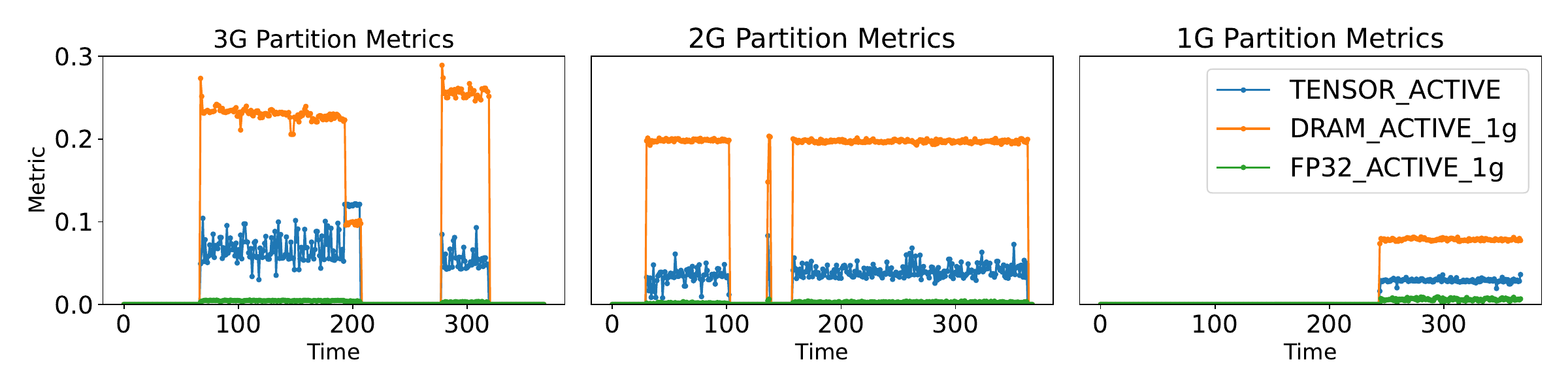}}
        \vspace{-0.1cm}
    \caption{\small DCGM metrics for concurrent executions of Llama, Granite, and Bloom  models on 3G, 2G, and 1G partitions, resp. (Legend shown in the third inset applies to the first two insets, as well.)}   
    \vspace{-0.4cm}
    \label{fig:MIG-1G-2G-3G-DCGM-metrics}
\end{figure}

\paragraph*{\hspace{-0.8 cm} Evaluation with 3 Concurrent Partitions}
We next demonstrate the scalability of MIG-features-based power estimation and partitioning by adding a 1G workload to the workload mix..

\begin{figure*}

        \hspace{-0.6cm}\includegraphics[scale=0.5]{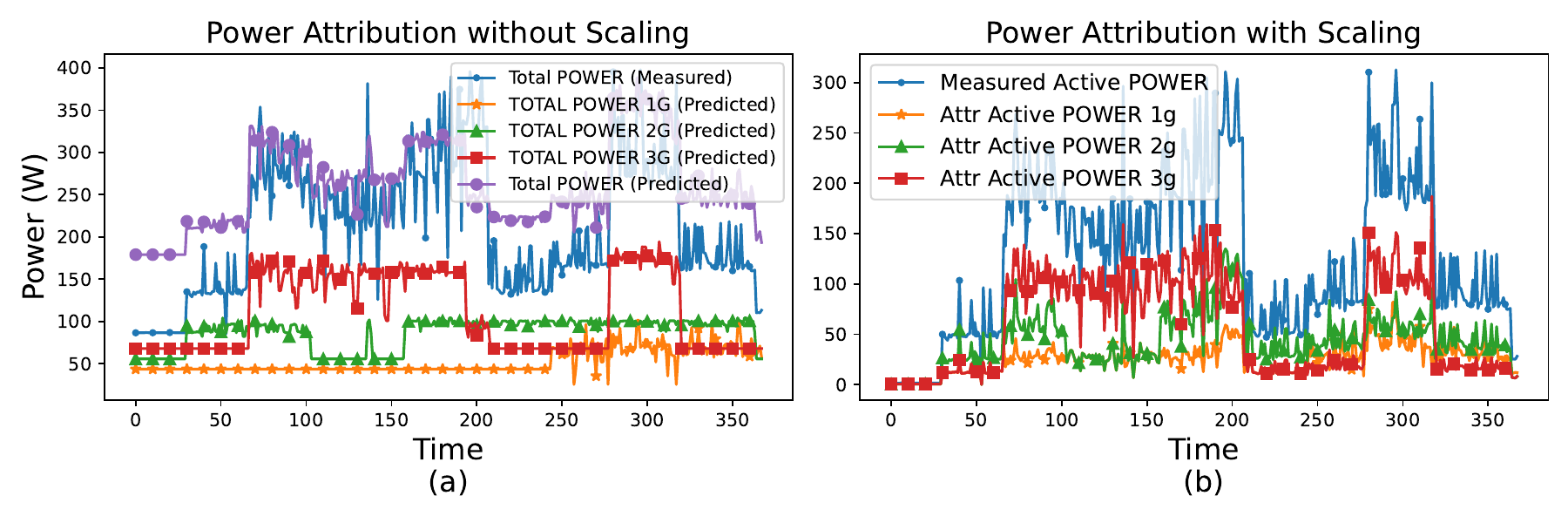}
    \vspace{-0.3cm}
    \caption{\small Power attribution to partitions via full-GPU models. {\bf(a)} Total measured and predicted partition powers without scaling {\bf(b)} Measured GPU active and attributed partition powers after scaling. }     
    \label{fig:MIG-1G-2G-3G-power-combo}
    \vspace{-0.35cm}
\end{figure*}

\begin{figure*}
     {\hspace{-0.8cm}   \includegraphics[scale=0.5]{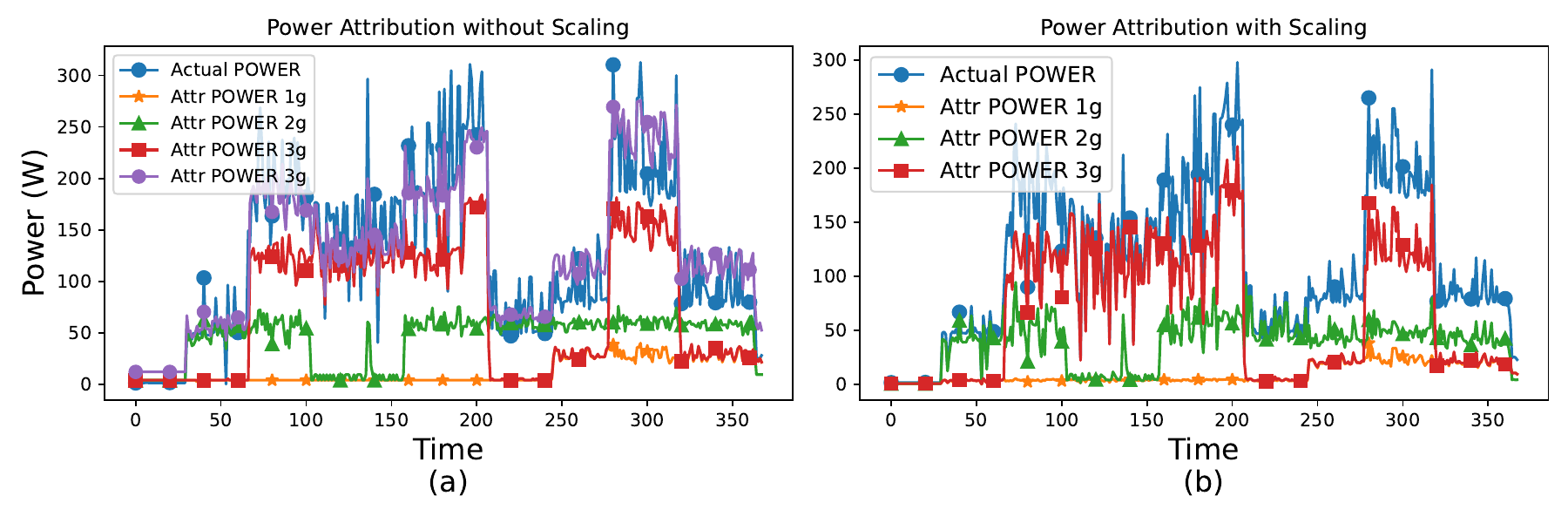}}
    \vspace{-0.3cm}
    \caption{\small Power attribution to partitions via runtime models based on MIG-specific features. {\bf(a)} Total measured and predicted partition powers without scaling {\bf(b)} Measured GPU active and attributed partition powers after scaling. }   
    \label{fig:MIG-1G-2G-3G-power-window-based-combo}
    \vspace{-0.35cm}
\end{figure*}

Fig.~\ref{fig:MIG-1G-2G-3G-DCGM-metrics} shows the DCGM metrics for the three workloads. Note that the 2G workload is the first to start at time $\sim$30, while 3G is loaded a little later at time $\sim$65 and 1G at $\sim$250. Note also that both 2G and 3G are stopped and resumed during the experiment to evaluate the impact of both these actions on power attribution to the remaining workloads.

Fig.~\ref{fig:MIG-1G-2G-3G-power-combo}(a) shows the plots of total GPU measured power and the total powers attributed to the three partitions as well as their sum with full GPU models, while Fig.~\ref{fig:MIG-1G-2G-3G-power-combo}(b) shows the measured GPU active power and the attributions of the same to the three partitions after scaling using the GPU power. Note that the total predicted power in the first case deviates significantly in intervals of lower utilization (e.g [200,270]), especially when the 3G partition is not loaded. This is rectified with scaling, but note that at time $\sim$65 when 3G is loaded, the attribution to 2G increases. However 3G's power does not change when 2G's load is reduced to 0 at time$\sim$100. Similarly, at time~280, resuming 3G causes power attribution of both 1G and 2G to increase.

Fig.~\ref{fig:MIG-1G-2G-3G-power-window-based-combo} plots the same numbers as Fig.~\ref{fig:MIG-1G-2G-3G-power-combo} but with a model based on MIG-specific features. Comparison of insets (a) of the figures shows the significant decrease to the error in the first-level predictions for MIG partitions. The residual error is eliminated via scaling. Comparison of the insets (b) of the two figures shows that unlike in the first case, there is not much perceptible change to 2G's power at times $\sim$65 and 280. Power attribution to 1G is also more stable in the second approach. These observations demonstrate that using MIG-specific features is a promising way to perform reliable and robust power estimation and distribution for MIG partitions although it is more expensive and cannot be done in a real-time manner. However, if power apportioning is required primarily for GHG emission reporting purposes, it can be done in an offline manner.


\section{Related Work}  \label{sec:rel}
\noindent 
Early efforts to model GPU power consumption focused on static V-F configurations, neglecting the impact of DVFS~\textcolor{red}{\cite{Hong2009,GPUWattch,Song2013,Nagasaka2010}}. Nagasaka et al.~\textcolor{red}{\cite{Nagasaka2010}} developed a power consumption model for a Tesla GPU (GTX285) using statistical methods to correlate hardware performance events with GPU power consumption, achieving an average prediction error of 4.7\%. However, the authors noted that this approach was ineffective for more recent GPUs, particularly those from the Fermi generation.
Hong et al.\textcolor{red}{\cite{Hong2009}} introduced a power model for a Tesla GPU (GTX280) by analyzing both the binary PTX and the device pipeline at runtime. This model achieves high accuracy in GPU power predictions through offline PTX analysis, but it is highly specific to the GPU, limiting its applicability across different architectures or varying core and memory configurations and is not applicable to runtime unknown workloads.
Song et al.\textcolor{red}{\cite{Song2013}} proposed a power model based on artificial neural networks using hard-to-get profiler-based metrics.
Leng et al.\textcolor{red}{\cite{GPUWattch}} incorporated Hong’s power model into the GPGPU-Sim\textcolor{red}{\cite{Bakhoda2009}} simulator, creating the GPUWattch tool. This tool supports only NVIDIA Tesla and Fermi GPU microarchitectures and estimates cycle-level GPU power consumption during application execution. Nonetheless, it assumes linear scaling of GPU domain power consumption with frequency, which has been shown to be inaccurate due to the non-linear behavior of voltage scaling in some GPUs\textcolor{red}{\cite{Mei2013,Guerreiro2014}}.
The authors in \textcolor{red}{\cite{goswami2010exploring}} utilized GPGPU-Sim to develop a performance model for GPGPU architecture that could potentially be extended to include a power model. However, this approach often necessitates modifications to the GPU scoreboard, making replication on actual hardware challenging.
Abe et al.\textcolor{red}{\cite{Abe2012}} argued that previous models are intricately product-specific and difficult to apply to modern GPUs, proposing DVFS-aware power regression models for NVIDIA Tesla, Fermi, and Kepler GPUs. 
These works highlighted the non-linear relationship between power consumption and voltage scaling, emphasizing the importance of considering these effects for accurate power modeling. 

Thus, while previous works have proposed methods to construct different types of GPU power models, there is no known work that addresses the issue of apportioning the aggregate GPU power among its MIG partitions.

\section{Conclusion and Future Work}
\label{sec:conc}
\noindent
This study makes significant progress in addressing the challenge of estimating power consumption for individual MIG instances within cloud data centers. By developing predictive models based on lightweight DCGM metrics, we provide a method for reliable power estimation without requiring additional hardware. Our findings reveal that advanced metrics with high computational overheads show promise of improving prediction accuracy, but being impossible to acquire, we rely on
%
simpler, normalized metrics to effectively estimate power consumption at the partition level. Our analysis also indicates that no single model is universally effective across all workloads and GPU architectures, underscoring the need for adaptable power estimation strategies using online data in cloud environments where efficient power management and accurate carbon emission reporting are essential.
Our future work will focus on automating the selection of the most appropriate predictive model based on the specific hardware and workload characteristics and also determining when the online model used for MIG power partitioning should be updated. This adaptive approach would enable more accurate and efficient power estimation across a wide range of GPU architectures and application scenarios.

\bibliographystyle{ieeetr}
\bibliography{references}

\end{document}